\documentclass[a4,11pt]{article}
\title{Energy benchmarks for water clusters and  ice structures \\ from an embedded
many-body expansion
}
\author{
M. J. Gillan$^{1,2,3}$, D. Alf\`{e}$^{1,2,3,4}$, P. J. Bygrave$^{5,6}$, C. R. Taylor$^{5}$ and F. R. Manby$^{5}$ \smallskip \\
$^1$London Centre for Nanotechnology, Gordon St., London WC1H 0AH, UK \smallskip \\
$^2$Department of Physics and Astronomy, University College London \\ Gower St., London WC1E 6BT, UK \smallskip \\
$^3$Thomas Young Centre, University College London \\ Gordon St., London WC1H 0AH, UK \smallskip \\
$^4$Department of Earth Sciences, University College London \\ Gower St., London WC1E 6BT, UK \smallskip \\
$^5$Centre for Computational Chemistry, School of Chemistry \\ University of Bristol, Bristol BS8 1TS, UK \smallskip \\
$^6$Department of Chemistry, University of Southampton \\ Highfield, Southampton SO17 1BJ, UK
}

\mathchardef\mhyphen="2D

\usepackage{a4wide}
\usepackage{graphicx}

\begin{document}
\maketitle
\abstract{
We show how an embedded many-body expansion (EMBE) can be used to calculate
accurate \emph{ab initio} energies of water clusters and ice structures using wavefunction-based
methods. We use the EMBE described recently by 
Bygrave \emph{et al.} (J. Chem. Phys. \textbf{137}, 164102 (2012)), in which
the terms in the expansion are obtained from calculations on monomers, dimers, etc. acted on by an
approximate representation of the embedding field due to all other molecules in the system, this field
being a sum of Coulomb and exchange-repulsion fields. Our strategy is to separate the total energy of
the system into Hartree-Fock and correlation parts, using the EMBE only for the correlation energy, with
the Hartree-Fock energy calculated using standard molecular quantum chemistry for clusters and plane-wave
methods for crystals. Our tests on a range of different water clusters up to the 16-mer show that
for the second-order M\o{}ller-Plesset (MP2) method the EMBE truncated at 2-body level reproduces
to better than $0.1$~m$E_{\rm h}$/monomer the correlation energy from standard methods. The use of
EMBE for computing coupled-cluster energies of clusters is also discussed. 
For the ice structures Ih, II and VIII, we find that
MP2 energies near the complete basis-set limit reproduce very well the experimental values of the 
absolute and relative binding energies, but that the use of coupled-cluster methods for many-body 
correlation (non-additive dispersion)
is essential for a full description. Possible future applications of the EMBE approach are suggested.
}

\section{Introduction}
\label{sec:intro}

Accurate energy benchmarks have long been important in calibrating methods for treating
the energetics of molecular systems (see e.g. Refs.~\cite{curtiss91,lynch03,jurecka06,karton11}). 
We are concerned here with benchmarks for water
systems, whose subtle hydrogen-bonding energetics has proved remarkably difficult
to characterize~\cite{schwegler04,sit05,santra08,schmidt09,wangj11,santra11,ma12}. 
Accurate energy benchmarks for small water clusters have often been used both
to parameterize force fields~\cite{burnham99,bukowski07,fanourgakis08,kumar10,wangy11,wangy09,babin12}
and to assess electronic-structure 
methods~\cite{santra08,anderson06,dahlke08,wang10,gillan12,medders13}. However, 
cooperative many-body effects are very important in 
water~\cite{ojamae94,xantheas94,lagutchenkov05,santra07}, and to
characterize them fully we need benchmarks for larger aggregates of molecules, including
condensed phases. We show here how recently developed embedding methods~\cite{bygrave12} enable
accurate energy benchmarks to be computed for large water clusters and ice structures
using correlated wavefunction-based methods.

The simplest wavefunction-based method for treating electron correlation is second-order
M\o{}ller-Plesset theory (MP2)~\cite{moller34,szabo82,helgaker00}, which by good 
fortune is already quite accurate for water.
The computational effort required by MP2 for a chosen basis set scales rapidly as
$N^5$ with number of molecules $N$, but nevertheless benchmark MP2 energies
within $\sim 0.1$~m$E_{\rm h}$/monomer of the complete basis-set (CBS) limit have been reported for clusters
of up to about $20$ molecules. Exploratory MP2 calculations on ice structures have also
been reported~\cite{erba09,hermann08}. The coupled-cluster technique at the CCSD(T) level (coupled
cluster with single and double excitations and perturbative triples)~\cite{szabo82,helgaker00} is considerably more accurate than
MP2, and is generally regarded as the ``gold standard'' for treating correlation. However, its challenging
$N^7$ scaling has so far made it difficult to achieve a high degree of basis-set convergence for anything
larger than the water hexamer~\cite{olson07,bates09}, though ambitious attempts to apply it to 
large water aggregates have been reported~\cite{yoo10}. 

We shall describe a method for computing MP2 and CCSD(T) benchmarks for large water
aggregates which employs an embedded version of the widely used many-body 
expansion (MBE)~\cite{xantheas94,hankins70,pedulla96}.
In the standard form of MBE, the total energy $E_{\rm tot} ( 1, 2, \ldots N )$ of a system of $N$ monomers
is expressed as:
\begin{equation}
E_{\rm tot} ( 1, 2, \ldots N ) = \sum_i E^{(1)} ( i ) + \sum_{i < j} E^{(2)} ( i, j ) +
\sum_{i < j < k} E^{(3)} ( i, j, k ) + \ldots \; .
\label{eqn:MBE}
\end{equation}
Here, the first term on the right is the sum of 1-body energies, where $E^{(1)} ( i )$ denotes the
distortion energy of monomer $i$ in the absence of all the other monomers as a function of the relative
coordinates specifying its geometry. Similarly, the second term is the sum of all the 2-body interaction energies,
where $E^{(2)} ( i, j )$ is the energy of the dimer consisting of monomers $i$ and $j$ in the absence of all the
other monomers minus the sum of 1-body energies $E^{(1)} ( i ) + E^{(1)} ( j )$. The higher-body terms
are defined in a similar way, as explained in detail in many previous papers. A popular 
strategy~\cite{bukowski07,wangy11,wangy09,babin12,medders13} for
exploiting MP2 or CCSD(T) benchmarks to create force fields for water has been to use the
benchmarks to create accurate parameterized representations of the low-order terms in the MBE, usually
up to 3-body terms, and then to use a model for the monomer multipole moments and polarizabilities for the
higher-body terms. Recent work~\cite{babin12,medders13} has made it clear that an accurate description of cooperative
effects represented by these higher-body terms is important.

The concept of ``embedded'' versions of MBE (we abbreviate to EMBE) has been discussed in a number
of previous papers~\cite{bygrave12,manby12,hirata05,leverentz09,wen12}.  These versions have
the same formal structure as eqn~(\ref{eqn:MBE}), but the $n$-body terms $E^{(n)}$ no longer refer to
the energies of clusters of $n$ monomers in free space, but instead to the energies of these clusters
embedded in an approximate representation of the potential due to all the other monomers in the system.
In some molecular systems, the electron density distribution on each monomer changes substantially
when the molecules form large aggregates. This is a strong effect in water~\cite{coulson66,silvestrelli99}, 
where the dipole moment
of the monomers increases from $1.86$~D in free space to $\sim 2.6$~D in ice, so that the interaction
between a pair of molecules is appreciably affected by the presence of their neighbors. In the standard
MBE, such effects are represented by higher-body terms in the series, but in EMBE the embedding potential
causes them to appear already in the 1- and 2-body terms. The EMBE that we use here is the one reported recently
by Bygrave \emph{et al.}~\cite{bygrave12}, in which the embedding potential is a sum of the Coulomb field due to the
electron densities of the monomers and a confining field arising from the Pauli repulsion due to these electron
densities. A summary of this EMBE will be given in Sec.~\ref{sec:techniques}. Our strategy in the present work
is to use the EMBE only for the correlation part of the energy, the Hartree-Fock part being computed accurately
using other methods, as explained below. The same general idea underlies the incremental correlation
method of Stoll and co-workers~\cite{stoll92,paulus06}, a version of which has already been shown to be
highly successful for the energetics of ice Ih~\cite{hermann08}.

We have two main aims in this work. First, we want to test the accuracy of EMBE truncated at 2-body level
for the computation of the MP2 correlation energy of water systems. We do this by comparing the
MP2 correlation energy of water clusters calculated directly by standard methods with values given by EMBE.
The clusters used for these comparisons range from the 6-mer to the 16-mer, and we study both 
equilibrium configurations and configurations drawn from random thermal samples. We shall show that
the 2-body-truncated EMBE appproximation is remarkably accurate for MP2. However, CCSD(T) corrections
to the correlation energy are not fully captured by truncation of the present form of EMBE at the 2-body level. 
The second aim of this work is
to use EMBE to find out whether MP2 near the CBS limit gives an accurate account of the energetics
of ice structures. We approach this question by computing the HF energy and MP2 correlation energy
of the ice Ih, II and VIII crystal structures and comparing their sum with the results of experiment
and benchmark quantum Monte Carlo (QMC) calculations. These ice structures are of great topical interest,
because it has recently been shown that standard DFT methods have difficulty in accounting 
for their relative energies~\cite{santra11}. We shall see that MP2 is surprisingly accurate for the energetics
of the chosen ice structures, but that a full description requires the inclusion of beyond-2-body correlation,
which is provided by CCSD(T).

\section{Techniques}
\label{sec:techniques}

The total energy $E_{\rm tot} ( 1, 2, \ldots N)$ of an assembly of $N$ monomers is the sum of its
Hartree-Fock energy and and its correlation energy:
\begin{equation}
E_{\rm tot} ( 1, 2, \ldots N ) = E_{\rm HF} ( 1, 2, \ldots N ) + E_{\rm corr} ( 1, 2, \ldots N ) \; .
\end{equation}
Our EMBE techniques for computing $E_{\rm corr}$ and the completely separate techniques used for
$E_{\rm HF}$ will be outlined in this Section.

\subsection{Correlation energy}
\label{sec:correlation}

Consider first the correlation energy of a single monomer $i$. In standard MBE, this would be the correlation
energy $E_{\rm corr}^{(1)} ( i )$ of the monomer in free space, with the collection of atomic positions specifying
the monomer geometry denoted by the symbol $i$. However, the electronic state of monomer $i$ is changed
by the presence of all the other monomers in the system, and the change can be approximately
represented by defining the correlation energy ${\tilde{E}}_{\rm corr}^{(1)} ( i )$
of monomer $i$ in a suitably defined field due to the other monomers. This field depends on the 
collection of atomic positions of all the $N$ monomers except for $i$. The total
embedded 1-body correlation energy is then the sum of all the ${\tilde{E}}_{\rm corr}^{(1)} ( i )$.
In the same way, we can define the embedded 2-body correlation energy
${\tilde{E}}_{\rm corr}^{(2)} ( i, j )$ of dimer $( i, j )$, which depends on the field produced by
all monomers except for $i$ and $j$. This is the total correlation energy
${\tilde{E}}_{\rm corr} ( i, j )$ of dimer $( i, j )$ in the field of all the other monomers minus
the embedded 1-body correlation energies of $i$ and $j$:
\begin{equation}
{\tilde{E}}_{\rm corr}^{(2)} ( i, j ) = {\tilde{E}}_{\rm corr} ( i, j ) -
{\tilde{E}}_{\rm corr}^{(1)} ( i ) - {\tilde{E}}_{\rm corr}^{(1)} ( j ) \; .
\end{equation}
By extension, we can define embedded 3-body correlation energies
${\tilde{E}}_{\rm corr}^{(3)} ( i, j, k )$ and higher-body correlation energies. The total
correlation energy of the $N$-monomer system is thus decomposed according to the identity:
\begin{equation}
E_{\rm corr} = \sum_i {\tilde{E}}_{\rm corr}^{(1)} ( i ) +
\sum_{i < j} {\tilde{E}}_{\rm corr}^{(2)} ( i, j ) +
\sum_{i < j < k} {\tilde{E}}_{\rm corr}^{(3)} ( i, j, k ) + \ldots \; \; .
\end{equation}
Note that the definition of the terms in this identity is completely analogous to standard MBE for correlation
energy, with the sole difference that the correlation energy of each $n$-mer is computed in the field
representing the influence of all the other monomers.

This EMBE is exact by construction, no matter what choice we make for the embedding potential, but
its convergence properties will be strongly affected by this choice. Here we use the form of embedding
potential due to Bygrave, Allan and Manby (BAM)~\cite{bygrave12}, which has been shown to work well for some molecular
crystals. We shall show that it works so well for the MP2 correlation energy of water systems that
the resulting EMBE can be truncated at the 2-body level without significant loss of accuracy. (We refer to
2-body-truncated EMBE in the following as EMBE-2.)

The BAM embedding potential is constructed using iterative Hartree-Fock calculations on the embedded monomers.
The iterative process is initiated by HF calculations on all $N$ monomers in their given geometry, each being 
treated as isolated in free space. This gives the HF ground-state electron density $\rho ( {\bf r} )$ for each monomer,
which is represented in a basis of Gaussian functions centred on the atomic sites.
The electron density of each monomer gives rise to a Coulomb field $V_{\rm coul}$ and a Pauli exchange-repulsion
field $V_{\rm rep}$. The Coulomb field produced by each monomer is simply the sum of the
fields due to the individual Gaussians. The approximation adopted for $V_{\rm rep}$ uses the fact~\cite{wheatley90}
that the exchange-repulsion energy between two molecules A and B can be quite accurately represented
as $k S_{\rm AB}$, where $S_{\rm AB}$ is the overlap integral of their electron densities $\rho_{\rm A} ( {\bf r} )$
and $\rho_{\rm B} ( {\bf r} )$:
\begin{equation}
S_{\rm AB} = \int d {\bf r} \, \rho_{\rm A} ( {\bf r} ) \rho_{\rm B} ( {\bf r} ) \; ,
\end{equation} 
and $k$ is a constant depending on the species involved. With this motivation, the exchange-repulsion
potential due to a monomer having density $\rho ( {\bf r} )$ is assumed to be 
simply $k \rho ( {\bf r} )$. The initial approximation
for the embedding field acting on any monomer $i$ is then the superposition of the fields $V_{\rm coul} + V_{\rm rep}$
coming from all the other monomers. The HF ground-state calculation on each monomer is now repeated,
but this time in the initial approximation for the embedding potential. This yields a new electron distribution
for each monomer, which is then used to recompute the potentials $V_{\rm coul}$ and $V_{\rm rep}$, and the
process is repeated to self-consistency within a specified tolerance. The embedding potential
constructed in this way is then used without further change in calculating the correlation energies
of the monomers, dimers, etc..., from which the energies of the EMBE are extracted.

We note the physical motivation underlying the BAM embedding potential. The substantial electron
redistribution in water and some other systems caused by assembling monomers into larger aggregates
can be expected to change the correlation energy. The intrinsically many-body nature of this redistribution
is accounted for by the self-consistent iterative procedure in BAM, which describes cooperative
effects due to both long-range Coulomb polarization and short-range exchange-repulsion. This gives
some reason to expect that the influence of cooperative electron redistribution on the correlation
energy may be adequately described by truncating EMBE at the 2-body level.

The BAM form of EMBE that we have outlined can be applied to both molecular clusters and molecular crystals, and we
shall present the results of both types of calculation, performed using a development version of
the {\sc molpro} code~\cite{molpro10,molpro12}. The calculations on 
crystals require the use of periodic boundary conditions,
and Ewald techniques are used to handle the long-range Coulomb parts of the embedding potentials,
as described in Ref.~\cite{bygrave12}. For periodic systems, the embedded 2-body correlation terms involving any
given monomer in the primary cell should in principle be summed over all monomers and their images in
all cells, but in practice we set a spatial cut-off radius $R_c$ beyond which correlation terms are neglected. 
Since the correlation energies are expected to fall off with distance $R$ as $1 / R^6$, extrapolation to the
$R_c \rightarrow \infty$ limit is straightforward, as we show later. Both the calculation of the self-consistent
embedding potentials and the computation of the 1- and 2-body embedded correlation energies
can readily be distributed over parallel processors, and it is efficient to do so.


\subsection{Hartree-Fock energy}
\label{sec:HF}

The accurate computation of $E_{\rm HF}$ for small and moderate clusters, performed here using
{\sc molpro}~\cite{molpro10,molpro12}, is completely standard, and needs 
no further comment. The computation of the Hartree-Fock energy of crystals
has a long history~\cite{pisani80,pisani96,paier05,gillan08,guidon09,paier09}, 
but basis-set convergence of $E_{\rm HF}$ to the high tolerance of $\sim 0.2$~m$E_{\rm h}$
$\simeq 5$~meV/monomer that we attempt to achieve here for ice structures is still not 
routine, and we summarise here the plane-wave techniques that we have used. In
Sec.~\ref{sec:clusters_ice}, we will report comparisons between {\sc molpro} and
plane-wave calculations of the HF binding energies of water clusters, which demonstrate
that the plane-wave techniques do, indeed, achieve the required accuracy.

Our plane-wave calculations of HF energy all employ the PAW (Projector Augmented Wave) 
technique~\cite{blochl94,kresse99}
implemented in the {\sc vasp} code~\cite{kresse96}. The underlying theory is outlined in Ref.~\cite{paier05}. As usual
when applying plane-wave techniques to isolated molecules and clusters, the calculations
are actually performed on the system in a large periodically repeated cell, whose volume
$\Omega$ is systematically increased until convergence is achieved. It is explained in Ref.~\cite{paier05} that 
when this approach is used for neutral molecules or clusters there are two dominant
contributions to cell-size error, both of which fall off as $1 / \Omega$. The first is due to
the dipole moment of the repeated system, and the second is a completely different
contribution arising from the treatment of exchange in periodic boundary conditions. The
two contributions can, if necessary, be separated using $k$-point sampling techniques~\cite{monkhorst76},
but in practice we simply recalculate the energy for a sequence of cubic repeated cells of
increasing cell-length $L$ using $\Gamma$-point sampling. We will show in Sec.~\ref{sec:clusters_ice}
that convergence of the total energy of the clusters of interest with respect to $L$ to within
$\sim 1$~meV or better can readily be achieved. However, even with perfect convergence with respect to $L$,
the HF energy with PAW will not be identical to that computed with {\sc molpro} at the CBS limit,
for two reasons. First, the PAW calculations do not include relaxation of the core orbitals (the O(1s) orbitals
in the present case). Second, the PAW treatment is not exact for the valence electrons in the core regions,
since it uses only a finite number of projectors. However, our comparisons for clusters will demonstrate
that the resulting errors are well within our specified tolerance.

For the ice structures, we use exactly the same PAW techniques for HF energy, and in this
case it is essential to achieve convergence with respect to $k$-point sampling. With the
standard Monkhorst-Pack sampling~\cite{monkhorst76}, the error due to insufficient $k$-point sampling falls off
as $1 / N_k$, where $N_k$ is the number of $k$-points in the full Brillouin zone. If we compute
HF energies for a sequence of increasing $N_k$ values, we can then extrapolate to the
$N_k \rightarrow \infty$ limit, and we find that convergence to this limit within better than $1$~meV/monomer
is straightforward to achieve.


\section{Water clusters and ice structures}
\label{sec:clusters_ice}

We will start our tests of EMBE by studying the binding energies of four isomers
of the hexamer (H$_2$O)$_6$, which have been accurately characterized in many previous 
papers~\cite{santra08,dahlke08,olson07,bates09}, and for
which highly converged HF and MP2 energies are readily computed by standard techniques. The
hexamers will also allow us to test our plane-wave techniques for computing the HF energy, which we rely
on later for the ice structures. We then present similar tests on equilibrium configurations of
the (H$_2$O)$_8$, (H$_2$O)$_{12}$ and (H$_2$O)$_{16}$ clusters. Further tests on samples of non-equilibrium
congurations of (H$_2$O)$_6$ and (H$_2$O)$_9$ follow. Having quantified the errors of EMBE, we will then 
turn to the energetics of ice Ih, II and VIII structures, where we can assess the accuracy of 
our HF and MP2 methods by comparing with energies from experiment and quantum Monte Carlo.

\subsection{Four isomers of the hexamer}
\label{sec:hexamer}

\begin{figure}[!htb]
\begin{tabular}{cc}
\includegraphics[width=0.5\linewidth]{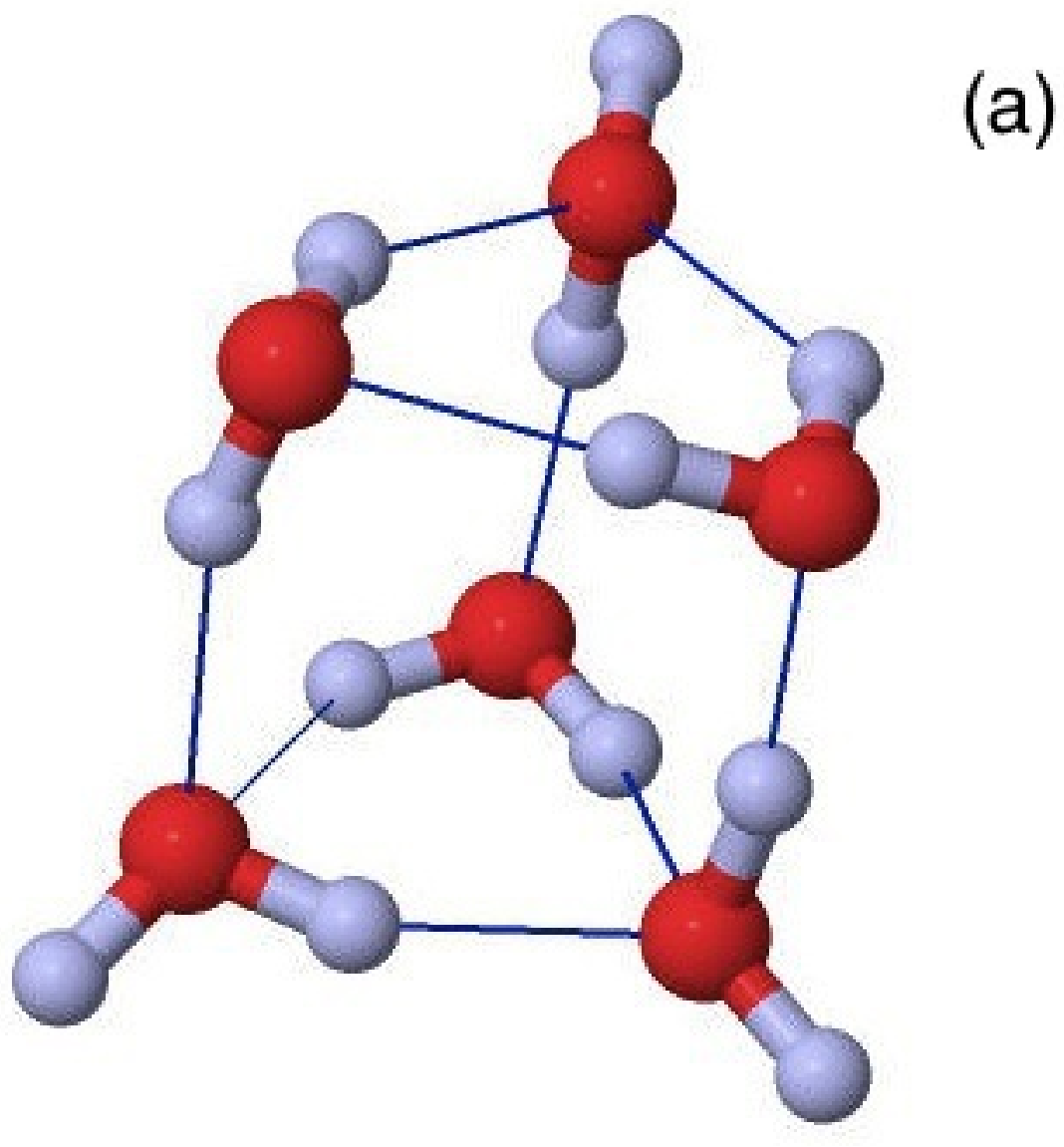} &
\includegraphics[width=0.5\linewidth]{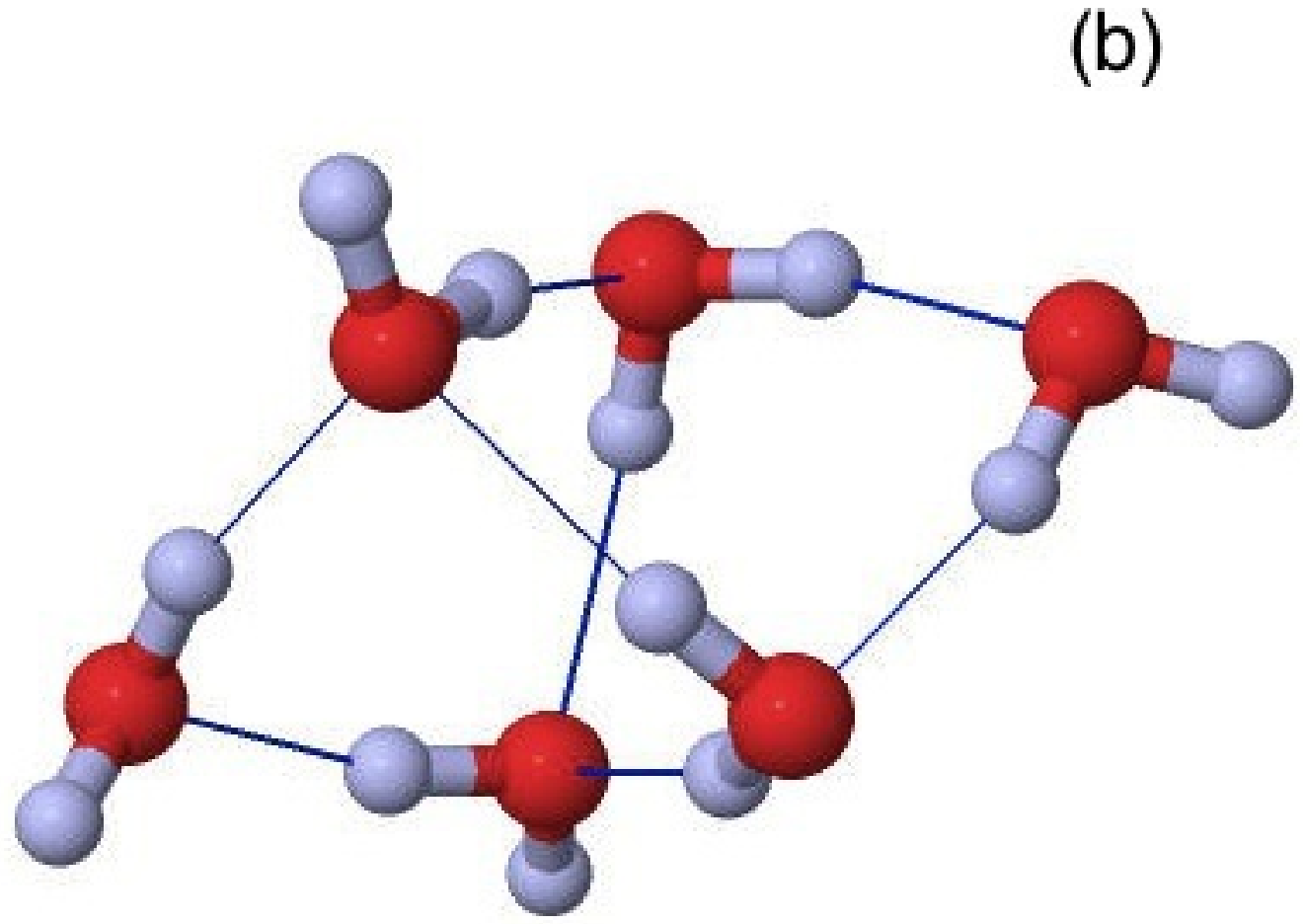} \\
\includegraphics[width=0.5\linewidth]{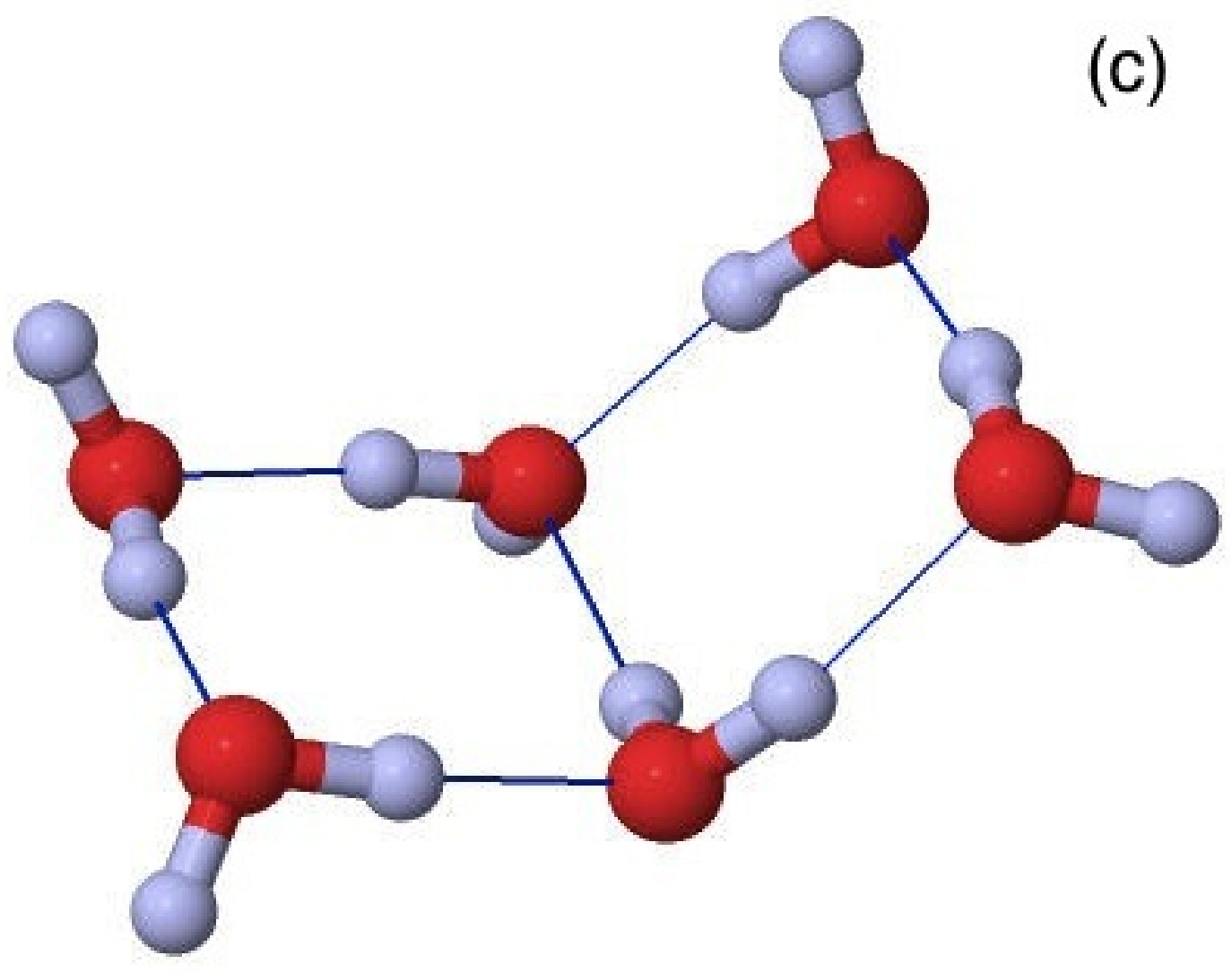} &
\includegraphics[width=0.5\linewidth]{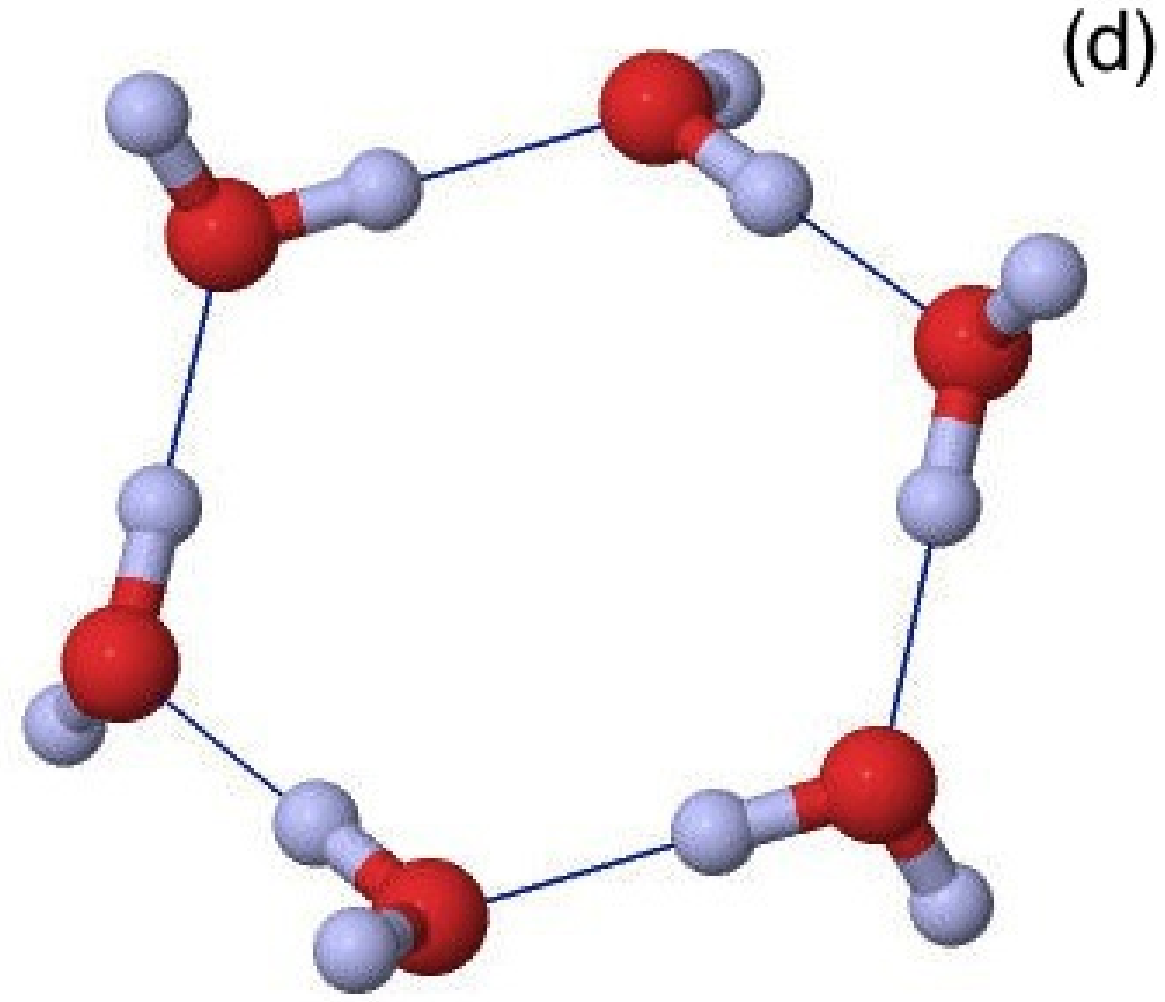}
\end{tabular}
\caption{
The four isomers of the H$_2$O hexamer whose binding energies we compute using
MP2 and CCSD(T): (a)~prism; (b)~cage; (c)~book; (d)~ring. Red and grey spheres represent
O and H atoms, with connecting lines showing hydrogen bonds.
}
\label{fig:hexamers}
\end{figure}

We compute the binding energies of the prism, cage, book and ring isomers of the 6-mer,
shown in Fig.~\ref{fig:hexamers}. The atomic
coordinates used for these calculations are taken from the work of Santra \emph{et al.} (Supplementary 
Information)~\cite{santra08}, who obtained them by relaxing the 
structures with MP2 using aug-cc-pVTZ basis sets. The binding energies
of the clusters and ice structures treated here are always referred to the energy of the appropriate
number of isolated H$_2$O monomers in the equilibrium geometry
reported in the definitive work by Partridge and Schwenke~\cite{partridge97}. For all our cluster calculations,
we use the correlation-consistent aug-cc-pVXZ basis sets~\cite{dunning89,kendall92} 
(we refer to them simply as AVXZ, with X 
the cardinality), approaching the CBS limit in the usual way by
increasing the cardinality. All our correlation energies are computed using the explicitly-correlated
F12 technique~\cite{werner07,adler07} provided by {\sc molpro}. All our calculations on embedded
dimers employ counterpoise. According to the tests reported in
Ref.~\cite{gillan12}, these technical choices are expected to deliver embedded 2-body correlation
energies to within better than $20$~$\mu E_{\rm h}$ of the CBS limit.

We report in Table~\ref{tab:hexamer} the directly calculated HF and MP2-correlation components 
$E ( \mathrm{HF} )$ and $E ( \Delta \mathrm{MP2(direct)} )$  of
the total binding energy of the four isomers, and also the MP2 correlation energies $E ( \Delta \mathrm{MP2(EMBE\mhyphen 2)} )$
given by EMBE-2. (We use the notation $\Delta \mathrm{MP2}$ to refer to the correlation energy computed with MP2.)
Comparing the $\Delta$MP2(direct) and $\Delta$MP2(EMBE-2)
values, we see that the errors incurred by truncating EMBE are very small, the worst error in the
total binding energy being $0.3$~m$E_{\rm h}$, which equates to $50$~$\mu E_{\rm h}$ ($1.4$~meV)
per monomer. We also compare the total MP2 binding energies relative to that of the prism
with the corresponding benchmark MP2 values reported recently in Ref.~\cite{gillan12}. This is an interesting
comparison, because it is well known that standard DFT approximations give erroneous trends in binding
energy as we pass from the compact prism and cage structures to the extended 
book and ring structures~\cite{santra08}.
The results of Table~\ref{tab:hexamer} show that the EMBE errors are small compared with the variation
of binding energy through the isomer sequence.

It is well known that the more accurate CCSD(T) treatment of correlation does not change the energy ordering
of the (H$_2$O)$_6$ isomers, but increases the difference of total energy between the prism and the ring
from $1.9$ to $2.9$~m$E_{\rm h}$~\cite{gillan12,bates09}. To test whether EMBE correctly predicts
this change, we used it to compare the total binding energies with CCSD(T) and MP2, using AVTZ basis sets
and F12. The AVTZ basis set is smaller than the AVQZ basis used for the MP2 calculations reported above,
but is expected to suffice for the difference 
$E ( \delta \mathrm{CCSD(T)} ) \equiv E ( \mathrm{CCSD(T)} ) -E ( \mathrm{MP2} )$.
Table~\ref{tab:hexamer} compares the $E ( \delta \mathrm{CCSD(T)} )$ values from EMBE with those from
standard methods~\cite{gillan12}, and we see that EMBE gives good results for the more extended ring and book
structures, but gives a somewhat overbinding $\delta \mathrm{CCSD(T)}$ shift for the compact cage and prism, so
that the increase of prism-ring splitting due to $\delta \mathrm{CCSD(T)}$ is exaggerated. An important
difference between MP2 and CCSD(T) is that the latter includes 3-body correlations, which are not accounted
for by 2-body-truncated EMBE, and this is the likely cause of the errors. (To keep the matter in perspective, though,
the worst of these errors is still less than $0.1$~m$E_{\rm h}$ per monomer.)

In order to test the plane-wave techniques used later to compute the HF energies of the ice structures,
we have calculated the HF binding energies of the (H$_2$O)$_6$ isomers using PAW and treating
the hexamer in a large periodically repeated cell, as described in Sec.~\ref{sec:HF}.
The cell length used was $30$~\AA, and we have checked that this is large enough to reduce residual cell-size
errors to less than $0.1$~m$E_{\rm h}$ in the total energy. The discrepancy between the PAW total binding energies
from {\sc vasp} (Table~\ref{tab:hexamer}) and the {\sc molpro} values is at worst $0.16$~m$E_{\rm h}$, or 
$27$~$\mu E_{\rm h}$ ($0.7$~meV) per monomer, which is negligible for present purposes.

\subsection{The octamer, dodecamer and hexadecamer}
\label{sec:8-12-16}

\begin{figure}[!htb]
\begin{tabular}{cc}
\includegraphics[width=0.4\linewidth]{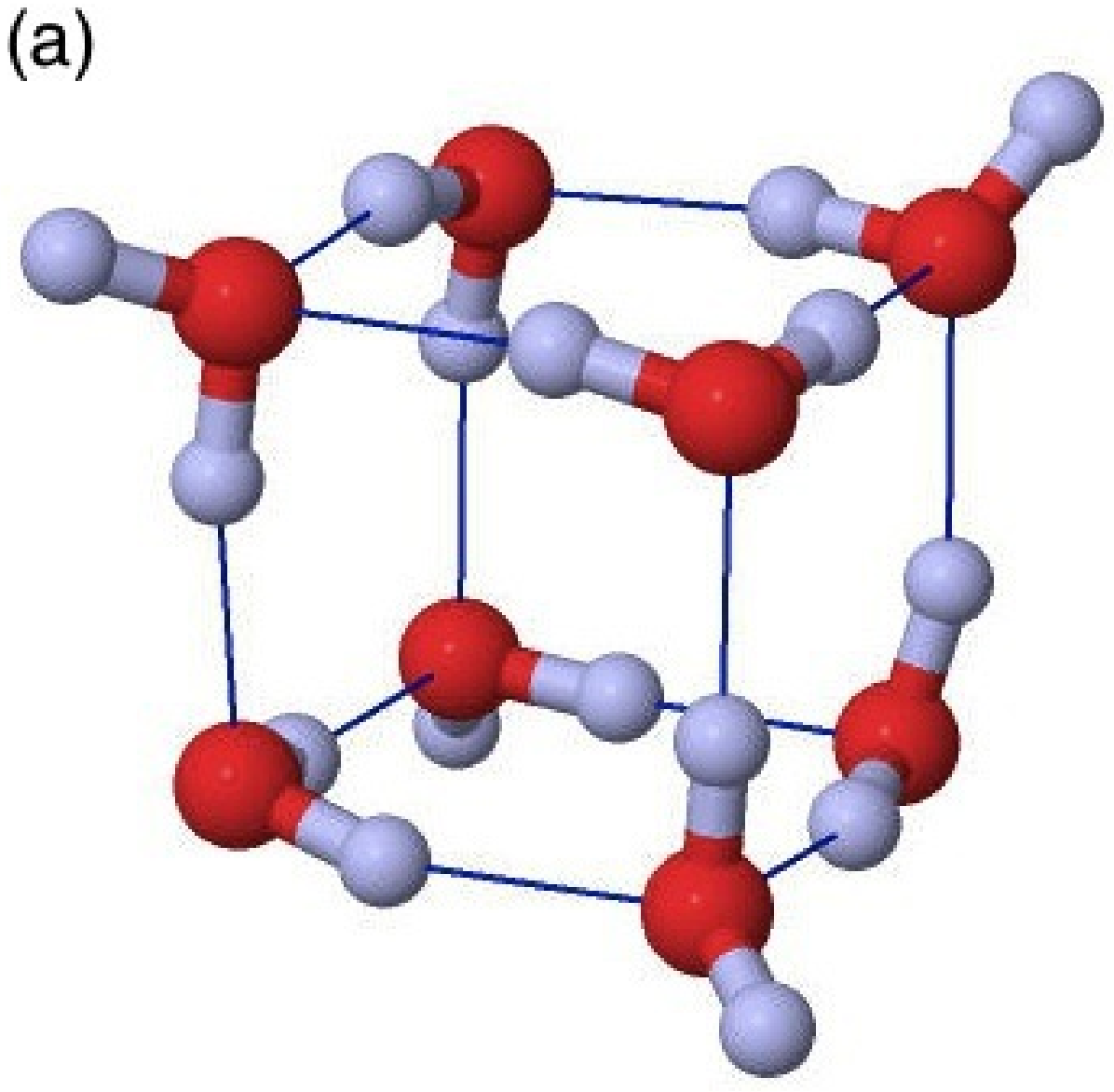} &
\includegraphics[width=0.5\linewidth]{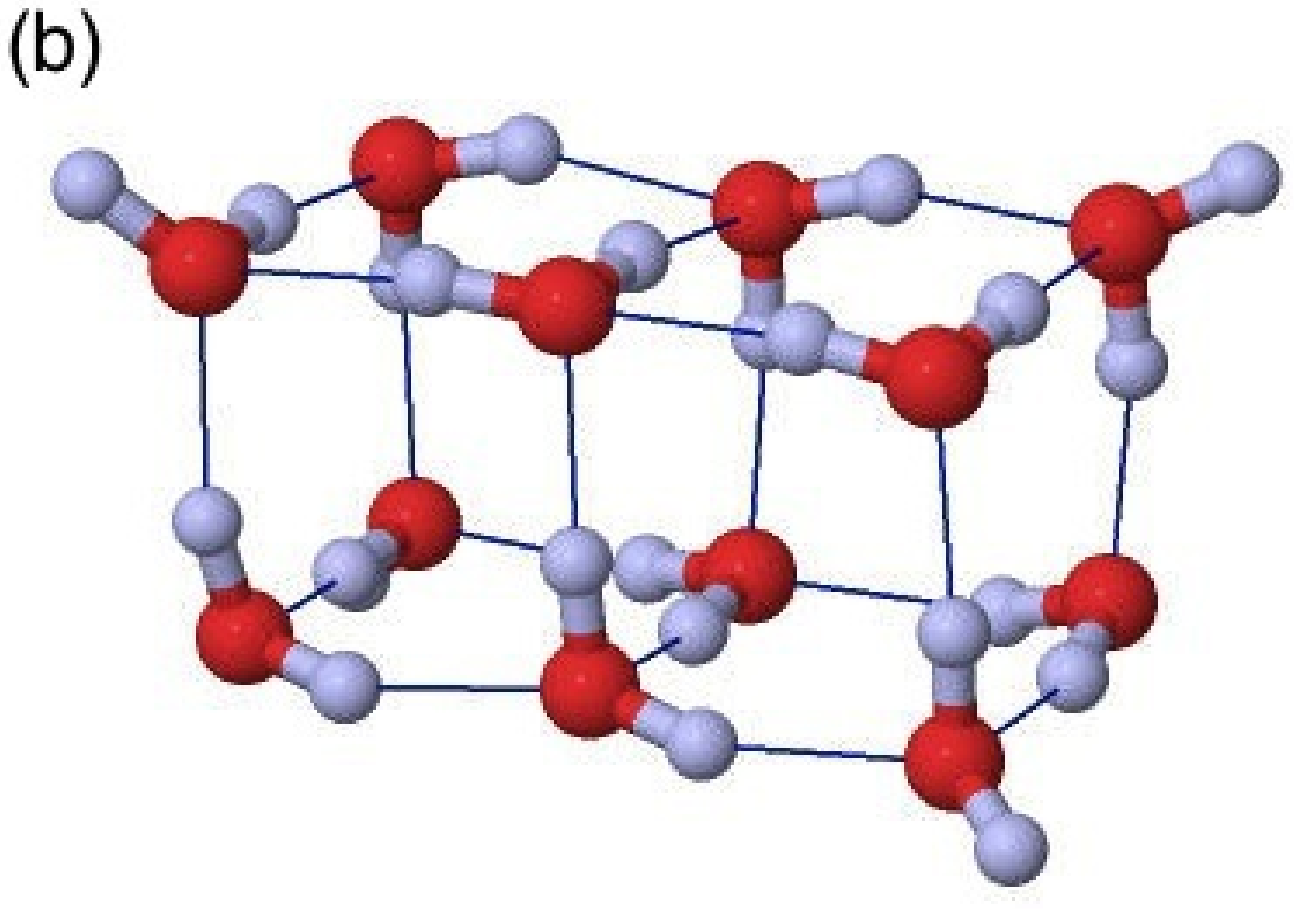} \\
\multicolumn{2}{c}{
\includegraphics[width=0.7\linewidth]{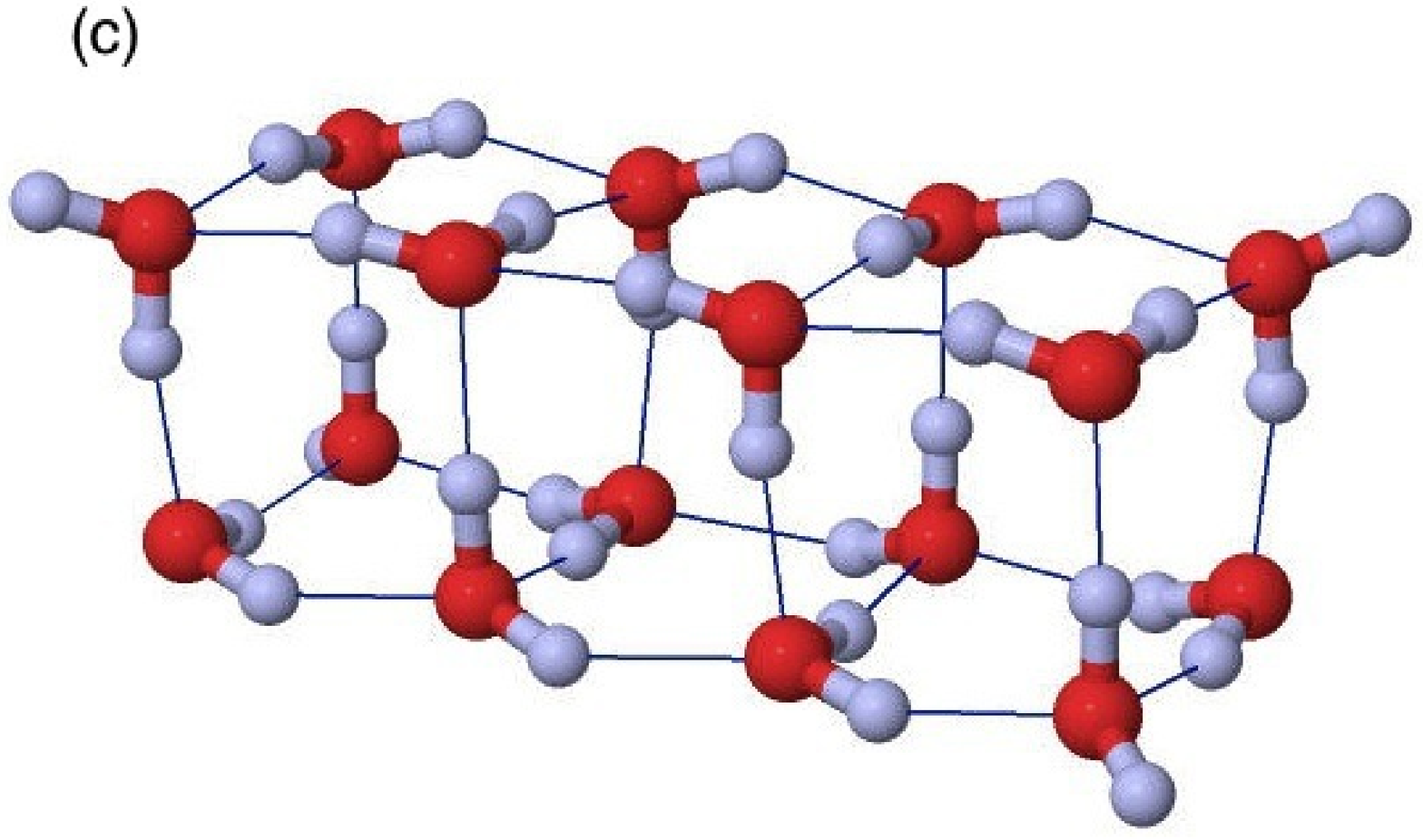}
}
\end{tabular}
\caption{
Geometries of the octamer (a), dodecamer (b), and hexadecamer (c) water clusters whose
binding energies we compute using MP2. Connecting lines show hydrogen bonds.
}
\label{fig:8-12-16mer}
\end{figure}

The geometries we have used for our calculations on the 8-mer, 12-mer and 16-mer all have the fused-cube
form, and are shown in Fig.~\ref{fig:8-12-16mer}. The atomic coordinates of 
the 8-mer and 12-mer were taken from the work
of Wales and Hodges~\cite{wales98}, in which the basin-hopping algorithm~\cite{wales03} was used together with the
empirical TIP4P interaction model~\cite{jorgensen83} to search for minimum-energy structures of a range of water clusters.
The geometry used here for the 16-mer was the one obtained  by Yoo \emph{et al.}~\cite{yoo10}
by energy minimization at the MP2/AVTZ level, and is the same as the geometry used in the recent
works by G\'{o}ra \emph{et al.}~\cite{gora11} and Wang \emph{et al.}~\cite{wang13}.

We report in Table~\ref{tab:8-12-16mer} the total MP2 binding energies of the clusters, and the HF
and MP2 correlation components of these binding energies. As before, the HF energies are computed
both using standard {\sc molpro} calculations with AVQZ basis sets and using PAW with the {\sc vasp} code.
We give values
for the correlation energies calculated with MP2-F12 and AVQZ basis sets both directly and using
EMBE-2. For comparison, we show also the MP2 binding energies for the 16-mer
reported recently in Ref.~\cite{wang13}. (The reference geometry of the free H$_2$O monomer used in
Ref.~\cite{wang13} differs slightly from the Partridge-Schwenke reference that we use, and we 
have adjusted their energies accordingly.)
As for the hexamers, the HF energies from PAW agree closely with those from {\sc molpro}, the largest
discrepancy being $0.5$~m$E_{\rm h}$ in the total energy for the 12-mer, which corresponds to 
$50$~$\mu E_{\rm h} \simeq 1.5$~meV per monomer. (The HF binding energy from Ref.~\cite{wang13}
given in the Table is also in close agreement.) The high accuracy of EMBE-2 for
the correlation component $E ( \Delta \mathrm{MP2} )$ of the MP2 energy is shown by the very close
agreement between the direct and EMBE-2 values for the octamer. It is further supported by the agreement
with the total MP2 binding energy of the hexadecamer reported by Wang \emph{et al.}~\cite{wang13}. (The
difference from their binding energy is perhaps slightly great than expected, but is less than $0.1$~m$E_{\rm h}$
per monomer.)

\subsection{Thermal samples of the hexamer and nonamer}
\label{sec:thermal-6-9}

The tests of EMBE that we presented so far all refer to equilibrium structures. However,  for many applications
such structures may not be particularly relevant, and it is therefore interesting to test EMBE for more general
configuration samples. In our recent work on benchmarking with QMC~\cite{gillan12,alfe13}, 
we have shown that it is useful
to work with random samples of configurations typical of thermal equilibrium. One type of system
that we studied was water ``nano-droplets'', i.e. clusters in thermal equilibrium, with evaporation prevented
by a weak confining potential. We present here tests of EMBE for 20 configurations each of the hexamer
and the nonamer in thermal equilibrium at a temperature of $200$~K. The techniques used to generate
these thermal-equilibrium samples are described in detail in Ref.~\cite{alfe13}.

\begin{figure}[!htb]
\begin{tabular}{cc}
\includegraphics[width=0.5\linewidth]{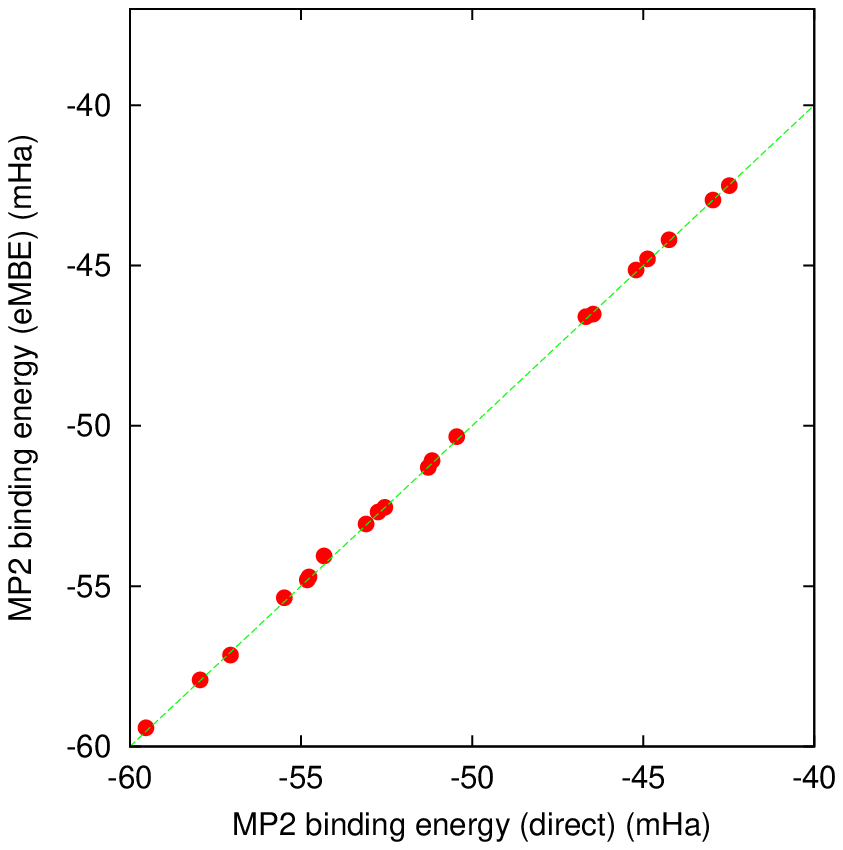} &
\includegraphics[width=0.5\linewidth]{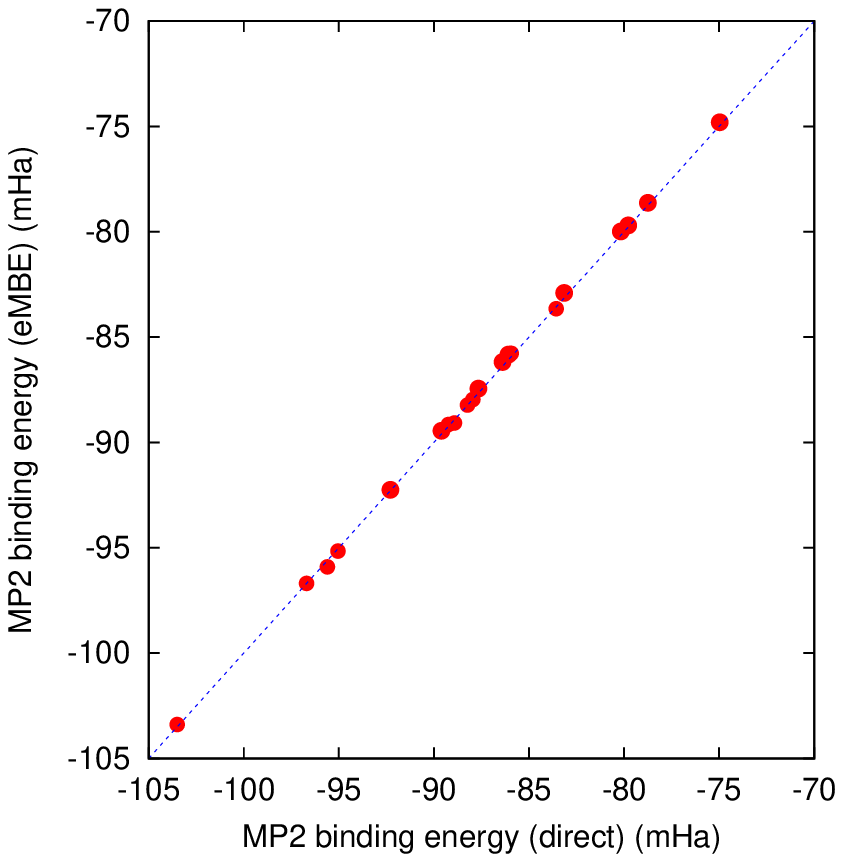}
\end{tabular}
\caption{
Parity plots characterizing accuracy of EMBE for MP2 binding energy of random
thermal samples of 20 configurations each of the H$_2$O hexamer (left panel)
and nonamer (right panel). Horizontal and vertical axes show total
binding energy (m$E_{\rm h}$ units) relative to free monomers computed using standard MP2 and using
MP2 with EMBE truncated at 2-body level.
}
\label{fig:thermal_parity}
\end{figure}

For small nano-droplets in thermal equilibrium, even at a temperatures as low as $200$~K, the binding
energy spontaneously fluctuates over a wide range, and we want to know whether EMBE makes
significant errors in describing these fluctuations. Using exactly the same techniques as for the other
clusters, we calculate the HF energy and the MP2 correlation energy directly for all the
configurations, using AVQZ basis sets, with F12 used for $\Delta$MP2. Separately, we use
EMBE-2 to compute $E ( \Delta \mathrm{MP2} )$, again with AVQZ and F12. Fig.~\ref{fig:thermal_parity} shows
parity plots of the total $E ( \mathrm{MP2} ) \equiv E ( \mathrm{HF} ) + E ( \Delta \mathrm{MP2} )$ 
binding energies computed in these two ways
for the 6-mer and the 9-mer. The errors of EMBE are so small that they are almost imperceptible on these
plots. The mean values of the EMBE errors, i.e. the mean deviations of MP2(EMBE) from
MP2(direct) are $50$ and $70$~$\mu E_{\rm h}$ in the total binding energy. The rms values
of these deviations of total energy are $75$ and $140$~$\mu E_{\rm h}$.

The overall conclusion from all the foregoing tests on clusters is that the errors incurred by using 2-body-truncated EMBE
for computing the MP2 correlation energy are no more than $100$~$\mu E_{\rm h}$ ($\simeq 3$~meV) per
monomer, which is negligible for most practical purposes.

\subsection{Ice structures}.
\label{sec:ice}

The normal form of ice at ambient conditions is the ice Ih structure~\cite{petrenko99}, 
in which each H$_2$O monomer
forms hydrogen bonds with four first neighbors at O-O separations of $2.7$~\AA, the second neighbors
having the much longer separation of $4.5$~\AA. Ice Ih is proton-disordered but is closely related to the
proton-ordered ice XI structure, which is the stable low-pressure form at 
temperatures below $72$~K~\cite{petrenko99}. With increasing
pressure at low temperatures, ice transforms successively to the sequence of denser
structures known as II, XV and VIII, the density of ice VIII at the pressure at which it first 
becomes stable being $\sim 70$~\% greater than that
of ice Ih (or ice XI)~\cite{petrenko99}. The large increase of density 
through this sequence results entirely from shortening
of the second-neighbor O-O distance, and in ice VIII each monomer is surrounded by eight neighbors
at almost equal O-O separations of $\sim 2.8$~\AA. Since four of these neighbors are not hydrogen-bonded
to the central monomer, Coulomb interactions and exchange-repulsion should strongly destabilize ice VIII
relative to ice Ih, but electron correlation will have the opposite effect, so that the relative energies
of ice VIII and ice Ih will depend on a balance between Coulomb, exchange-repulsion and correlation 
energies. It is known that standard DFT approximations describe this balance rather poorly~\cite{santra11}, making the
$\mathrm{VIII} - \mathrm{Ih}$ energy difference much too great. The question we address here is whether MP2 describes this balance correctly.

\begin{figure}[htb]
\centering
\includegraphics[width=0.8\linewidth]{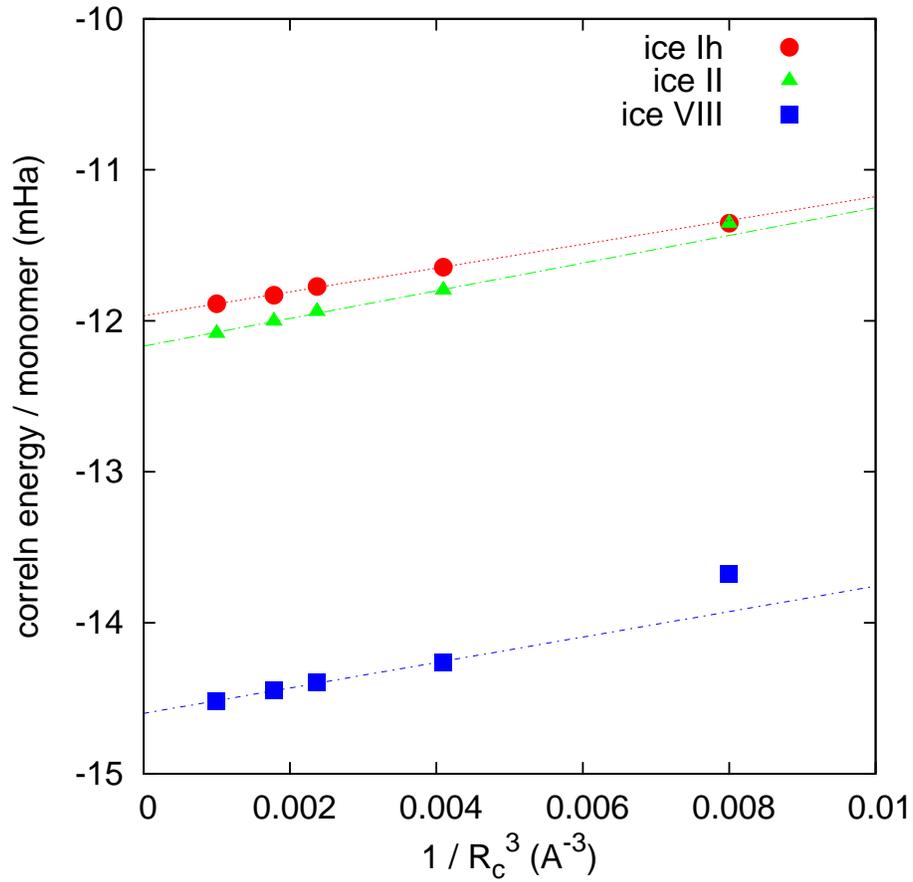}
\caption{
The correlation part of the binding energy per monomer for the ice structures Ih, II and VIII computed with
the embedded many-body expansion truncated at 2-body level as function of cut-off radius $R_c$ (see text).
Correlation energy is computed with MP2-F12 using AVQZ basis-sets. Straight lines are least-squares fits
to the points having $R_c \ge 6.25$~\AA\ (i.e. omitting the point at $1 / R_c^3 = 0.008$~\AA).
}
\label{fig:ice_correl_extrap}
\end{figure}

The atomic coordinates for our ice calculations are exactly the same as those used in the
QMC calculations of Ref.~\cite{santra11}. We present 
first our EMBE calculations of correlation energy with MP2, for which we use AVQZ basis sets with F12.
We performed a thorough test of $k$-point convergence for the ice VIII structure, by monitoring the total
HF energy of the 8-molecule cell as the number of $k$-points was increased stepwise from $1$ to $1152$.
We found that for $75$ $k$-points or more this total energy was converged to better than $30$~$\mu E_{\rm h}$
($\sim 1$~meV). For all the results that follow, the number of $k$ points was always large 
enough to ensure this degree of convergence.
As noted in Sec.~\ref{sec:correlation}, the embedded 2-body
correlation energy includes all dimers within a specified cut-off radius $R_c$, and we must ensure that
the MP2 correlation energy $E ( \Delta \mathrm{MP2} )$ is adequately converged with 
respect to $R_c$. Since correlation at long distances $r$
is dispersion, which falls off as $1 / r^6$, we can estimate the error
$\delta E_{\rm corr}$  incurred by truncating the sum over
2-body correlation contributions at radius $R_c$ as:
\begin{equation}
\delta E_{\rm corr} = 2 \pi n_{\rm mol} \int_{R_c}^\infty dr \, r^2 C_6 / r^6 \; ,
\end{equation}
where $n_{\rm mol}$ is the number density of monomers and $C_6$ is the dispersion coefficient. The
error $\delta E_{\rm corr}$ therefore falls off as $1 / R_c^3$, so that we can estimate the $R_c \rightarrow \infty$
limit by plotting the correlation energy against $1 / R_c^3$ and making a linear extrapolation. Our
MP2 correlation energies for $R_c$ values ranging from $5$ to $10$~\AA\ are plotted in this way for the
three ice structures in Fig.~\ref{fig:ice_correl_extrap}, together with 
linear least-squares fits for $R_c \ge 6.25$~\AA, and we
see from this that the uncertainty due to $R_c \rightarrow \infty$ extrapolation is no more than $\sim 0.1$~m$E_{\rm h}$/monomer.

Our Hartree-Fock energies, together with the MP2 correlation energies obtained by $R_c \rightarrow \infty$
extrapolation, and the resulting MP2 cohesive energies are reported in Table~\ref{tab:ice}, where we also
give experimental values~\cite{whalley84} and values from QMC calculations~\cite{santra11}.
The agreement of MP2 with these accurate
benchmark values is surprisingly good, the deviations from experiment all being less than
$0.25$~m$E_{\rm h}$ ($7$~meV) per monomer and the largest difference  from the QMC energies
being $0.31$~m$E_{\rm h}$ ($9$~meV) per monomer.

It might be tempting to conclude from these comparisons that MP2 gives a complete and accurate
description of ice energetics. However, it would be rash to do so without assessing the corrections given by
the more accurate CCSD(T) approximation. We have followed the same procedure as for the
hexamers (Sec.~\ref{sec:hexamer}), obtaining the difference 
$E ( \delta \mathrm{CCSD(T)} ) \equiv E (  \mathrm{CCSD(T)} ) - E ( \mathrm{MP2} )$ between
the CCSD(T) and MP2 correlation energies by using EMBE-2 to compute the
CCSD(T) and MP2 binding cohesive energies with AVTZ basis sets and F12. Our calculations of the
difference $E ( \delta \mathrm{CCSD(T)} )$ were performed with cut-off distance $R_c$ equal to $7.5$~\AA,
which our tests indicate is large enough to ensure convergence to better than
$0.1$~m$E_{\rm h}$ per monomer. Our values of the $\delta \mathrm{CCSD(T)}$ corrections to the correlation
energy per monomer for the three ice structures (Table~\ref{tab:ice}) have the effect of stabilizing ice VIII and
destabilizing ice Ih, and are so sizeable that the cohesive energies of the two structures become almost
identical, so that the apparently excellent predictions of MP2 are completely overturned.

A likely explanation for this outcome is that truncation of EMBE at the 2-body level is not an accurate enough
 approximation for CCSD(T). In fact, our EMBE calculations on the hexamer (Sec.~\ref{sec:hexamer})
showed that although EMBE-2 gives satisfactory $\delta \mathrm{CCSD(T)}$ corrections
for the extended ring and book isomers, it gives significantly overbinding corrections for the compact
cage and prism. Since CCSD(T) includes beyond-2-body correlations but MP2 does not, EMBE-2 will
not be accurate for $\delta \mathrm{CCSD(T)}$ corrections if such correlations are important.

To investigate this further, we have assessed the significance of 3-body correlations by computing
$\delta \mathrm{CCSD(T)}$ corrections to the 3-body energies of the H$_2$O trimers that occur
in ice Ih and VIII. Our estimates are reported in the Appendix, where we show that the corrections
due to 3-body correlation are very small in ice Ih (see also Ref.~\cite{vonlilienfeld10}), but are
substantial in ice VIII, being large enough to destabilize the latter by $\sim 1.2$~m$E_{\rm h}$ per
monomer. This suffices to compensate almost entirely for the 2-body $\delta \mathrm{CCSD(T)}$
changes. We include in Table~\ref{tab:ice} our estimates for the 3-body $\delta \mathrm{CCSD(T)}$
corrections, and the resulting final CCSD(T) binding energies. The energy differences between ice VIII
and the other two ice structures are once more in reasonable agreement with experiment and QMC,
though we note that ice II has now become very slightly more stable than Ih. More work
will be needed to refine the details, but it seems clear that our understanding of ice energetics is incomplete
without 3-body correlation.


\section{Discussion and conclusions}
\label{sec:discussion}

A number of conclusions can be drawn from our results. First, the embedded many-body
expansion (EMBE) truncated at the 2-body level provides a remarkably effective way of calculating
the MP2 correlation energy of water systems. Our results imply that beyond-2-body terms in EMBE
contribute almost nothing to the MP2 correlation energy. 
This is important, because MP2 is widely used as a reasonably good approximation
to the energetics of water systems. A second important conclusion is that 2-body truncated
EMBE for MP2 correlation, when combined with accurate methods for the Hartree-Fock energy,
provides a rather straightforward way of computing the MP2 energies of ice structures close to
the basis set limit. The MP2 values for both the cohesive energies and the relative energies
of the Ih, II and VIII structures agree surprisingly well with accurate values from experiment
and from quantum Monte Carlo calculations. The results make it clear that the small energy differences
between the structures depend on a fine balance between Hartree-Fock and correlation energies, which
individually vary quite substantially. Specifically, the destabilization of ice VIII relative to Ih by
Coulombic and exchange-repulsion energies described by Hartree-Fock is to a large extent compensated
by the restabilization due to correlation. However, our EMBE calculations with CCSD(T) show that
the accuracy of MP2 for the relative energies of ice structures is not quite what it 
seems, and is partly due to a lucky cancellation
of errors in the description of 2-body and beyond-2-body correlation. Our analysis implies that
the energy difference between the Ih and VIII structures cannot be understood without
many-body dispersion. 

The success of the 2-body-truncated EMBE for MP2 correlation is not unexpected. Since MP2
includes only double excitations, it accounts for correlation between electrons on single monomers
and on pairs of monomers, but not for correlation between three or more monomers. In the standard
(unembedded) many-body expansion for correlation energy, beyond-2-body contributions arise from the
change of 2-body correlation energy due to polarization of monomers, and these effects are
accounted for by EMBE truncated at the 2-body level. The genuine 3-body correlation effects that we have seen
to be important in ice VIII are missed by MP2 but captured by CCSD(T). However, they are not captured
by the form of EMBE used in the present work if we truncate at the 2-body level. More sophisticated 
projector-embedding techniques (see e.g. Ref.~\cite{manby12}), which are expected to
yield still more rapidly convergent EMBE expansions, will be capable of including such many-body
correlation effects, even when truncated at 2-body level.

There is quite a close connection between EMBE as we have used it here and the incremental correlation
method pioneered by Stoll and co-workers~\cite{stoll92,paulus06}. In that method too, the total interaction
energy is separated into its HF and correlation parts, and a many-body expansion is used to compute
the correlation part. It was pointed out~\cite{hermann07} that the standard (unembedded) many-body
expansion applied to small water clusters converges more rapidly for the correlation energy than for the
total interaction energy, and this insight formed the basis for the demonstration by Hermann and
Schwerdtfeger~\cite{hermann08} that a 1- and 2-body treatment of the 
correlation energy computed using MP2 and CCSD(T)
gives an accurate account of the energetics of ice Ih. The EMBE approach is in principle more general,
since it allows for many different forms of embedding, and this flexibility may well be important
for future developments.

It seems likely that MP2 and CCSD(T) calculations in the EMBE approximation will be useful for predicting the
energetics of a range of other water systems. Examples might include the formation energies of
ordered and disordered ice surfaces~\cite{pan10} and the energetics of 
lattice defects in bulk ice~\cite{dekoning06}. Recent work
has demonstrated that 2-body-truncated EMBE can also be effective for other simple molecular
systems such as CO$_2$ and HF~\cite{bygrave12}. Beyond this, the energetics of mixtures should also be accessible,
a particularly important example being gas hydrates, including the environmentally important clathrates
of CH$_4$ and CO$_2$.

Two important technical features of EMBE will greatly assist its application to more complex systems.
The first is its very favourable computational scaling with system size. The effort needed for the computation of the
embedding potential is proportional to the number of monomers in the cluster or in the unit cell of the crystal. 
For clusters, the calculation of correlation scales as $N^2$, if we include correlation between all 
monomer pairs. For a crystal, on the other hand, once we have adopted a cut-off radius $R_c$ for 
the pair correlations, the scaling of the correlation calculation is linear in the number of monomers in the unit cell.
The example of gas hydrates is instructive here. The methane hydrates have typically $\sim 50$ molecules in the
unit cell, but the linear scaling of the correlation energy means that correlation for these systems requires
only $\sim 4$ times the computational effort needed by the ice structures treated here. This all means  that
if a standard method is used for the HF energy, then it is this that will dominate the overall scaling.
The second beneficial feature of EMBE is that it is trivially parallelizable. Since
the 2-body correlation energies are independent of each other, the wall-clock time needed for the total correlation
energy can in principle be reduced to the time needed for a single dimer, provided enough processors
are available.

Before concluding, we draw attention to the calculation of forces in the EMBE framework. In the present
work, we have calculated only energies, but clearly EMBE would become even more useful if it could
be used to relax structures and to perform molecular dynamics simulation. The calculation of forces is
an important problem for the future.

In conclusion, we have shown that a simple form of embedded many-body expansion truncated at 2-body level
gives an effective and accurate way of computing the MP2 correlation energy of large water clusters
and ice structures with rather modest computational resources. Our calculations on three key ice
structures show that MP2 gives a rather accurate account of their cohesive energies and their
relative energies. However, the success of MP2 for ice structures is partly due to error
cancellation, and a full understanding of their relative energies requires non-additive dispersion
as described by CCSD(T). The application of the embedding techniques to a range of other condensed-phase 
molecular systems appears likely to be both feasible and fruitful.

\section*{Acknowledgments}

CRT is supported by an EPSRC studentship. We thank Prof. K. D. Jordan for providing technical details of calculations on the water 16-mer performed
by his group.



\clearpage

\section*{Appendix: Three-body correlation in ice}

In Sec.~\ref{sec:ice}, we noted reasons for thinking that the relative energies of the ice Ih and VIII
structures cannot be fully understood without considering 3-body electron correlation (non-additive
dispersion). Here, we provide evidence confirming this idea. Since MP2 consists of second-order
perturbation theory starting from the HF ground state, it does not account for 3-body correlation.
By contrast, CCSD(T) does describe such correlation, since it includes contributions from all
orders of perturbation theory. The difference
$E^{(3)} ( \delta \mathrm{CCSD(T)} ) \equiv E^{(3)} ( \mathrm{CCSD(T)} ) - E^{(3)} ( \mathrm{MP2} )$
between 3-body energies calculated with CCSD(T) and MP2 provides some measure of 3-body correlation.

\begin{figure}[!htb]
\centering
\includegraphics[width=0.7\linewidth]{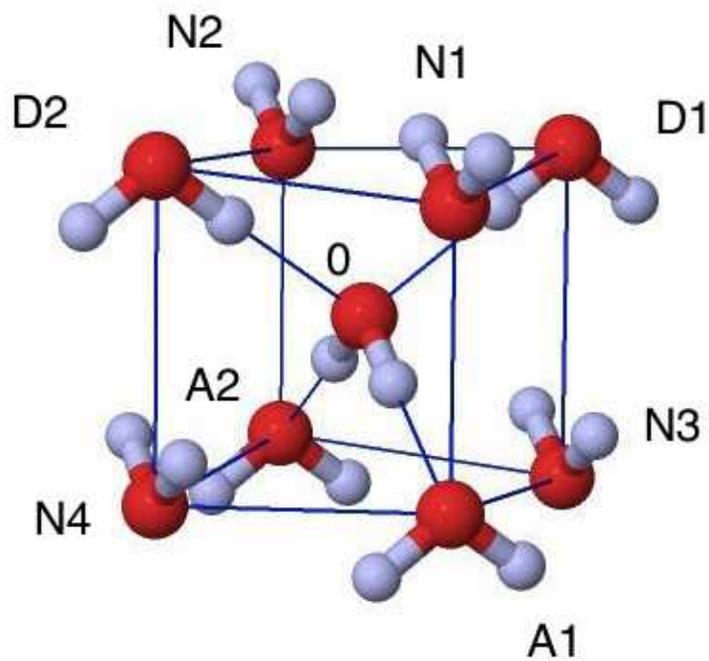}
\caption{
Water nonamer cut from ice VIII crystal structure. Labels indicate central monomer (0), donor (D1, D2)
and acceptor (A1, A2) hydrogen-bonded neighbors and non-bonded neighbors (N1 - N4). 
}
\label{fig:iceVIII_nonamer}
\end{figure}

We show in Fig.~\ref{fig:iceVIII_nonamer} the nonamer obtained by 
cutting out of the ice VIII crystal an H$_2$O monomer
and its eight nearest neighbors. Four of these neighbors are hydrogen bonded to the central
monomer as donors or acceptors (D1, D2, A1, A2 in the Figure, with central monomer labeled 0),
and the other four are non-bonded (N1 - N4 in the Figure).
The O-O distances from the central monomer to the eight neighbors
are all rather close to the O-O distance of $2.7$~\AA\ in ice Ih, so that the pentamer obtained by
taking the central H$_2$O and its four hydrogen-bonded neighbors is almost
the same as the pentamer that would be obtained by cutting out of the ice Ih crystal
a monomer and its four neighbors. There are 84 trimers that can be
formed by extracting all distinct triplets from the nonamer. We have computed
$E^{(3)} ( \delta \mathrm{CCSD(T)} )$ for all these trimers using AVDZ basis sets with F12 for both
MP2 and CCSD(T), and we use these $E^{(3)} ( \delta \mathrm{CCSD(T)} )$ values to assess
3-body correlation in ice. To reduce basis-set superposition error, we use full counterpoise,
so that the total energy and its 1- and 2-body parts are all computed using the basis set of the entire
trimer. Tests on a random sub-set of the 84 trimers show that repeating the calculations
with AVTZ basis sets produces differences of only a few $\mu E_{\rm h}$ for each trimer. 

It is helpful to separate the 84 trimers into the following groups (we give examples of group members
using the labeling of Fig.~\ref{fig:iceVIII_nonamer}):
\begin{itemize}
\item[G1:]
central monomer with one hydrogen-bonded neighbor and one non-bonded neighbor, the
two neighbors being adjacent to each other (example: D1-0-N1);
\item[G2:]
central monomer with two hydrogen-bonded neighbors (examples: D1-0-D2, D1-0-A1, A1-0-A2);
\item[G3:]
central monomer with two non-bonded neighbors (example: N1-0-N2);
\item[G4:]
central monomer with one hydrogen-bonded neighbor and one non-bonded neighbor, the
two neighbors being on opposite sides of the central monomer (example: A1-0-N2);
\item[G5:]
three neighbors of the central monomer, all lying on the same cube face (examples: D1-N1-A1, N1-D2-N2);
\item[G6:]
three neighbors of the central monomer, not all lying on the same cube face (example: A1-D1-D2).
\end{itemize}
The sums of all the $E^{(3)} ( \delta \mathrm{CCSD(T)} )$ values in each of these six groups are
reported in Table~\ref{tab:3Bcorr}. The contribution from group G1 dominates all the others, and groups G2 and G6 can
safely be ignored.

The energies in Table~\ref{tab:3Bcorr} can be used to estimate the contribution of 3-body correlation
per monomer in ice VIII. A little thought shows that for each monomer in a large sample of ice VIII
the number of trimers of type G1 is 12, which is the same as the number for the nonamer, so that we
can use the G1 entry in Table~\ref{tab:3Bcorr} as it stands. The same is true of all the other entries, except for
group G5, where a factor $1/2$ must be applied. Adding the contributions, we estimate that
$E^{(3)} ( \delta \mathrm{CCSD(T)} )$ raises the energy of ice VIII by $1.2$~m$E_{\rm h}$
($\simeq 32$~meV) per monomer. Table~\ref{tab:3Bcorr} indicates that for ice Ih the energy shift
due to $E^{(3)} ( \delta \mathrm{CCSD(T)} )$ is negligible, since the only contributions that need
to be considered in that structure are of type G2.

It is natural to ask whether $E^{(3)} ( \delta \mathrm{CCSD(T)} )$ can be understood in the framework of
the standard Axilrod-Teller-Muto (ATM) theory of 3-body dispersion~\cite{axilrod43,stone13}. According to ATM,
the 3-body dispersion interaction $E_{\rm ATM}$ of three identical, spherically symmetric bodies is:
\begin{equation}
E_{\rm ATM} = C_9 ( 1 + 3 \cos \gamma_1 \cos \gamma_2 \cos \gamma_3 ) / ( R_1 R_2 R_3 )^3 \; ,
\end{equation}
where $R_i$ and $\gamma_i$ are the sides and angles of the triangle formed by the three bodies,
and $C_9$ is a positive constant. To test whether this formula can account for the energies in Table~\ref{tab:3Bcorr},
we choose the value of $C_9$ so as to reproduce exactly the 3-body correlation energy of group G1, and we
then use this value to predict the contributions from the other groups. The required value of $C_9$ is
$289$~a.u., and the Table gives the resulting ATM energies, which agree very well with our
$E^{(3)} ( \delta \mathrm{CCSD(T)} )$ values. As predicted by the formula, most of the
correlation energies are positive, except for group G4, where one of the angles in the triangle
is $180^\circ$, so that the angular factor has the negative value $-2$. For groups G2 and G3,
the tetrahedral angle ($109.5^\circ$) is close to the value of $117^\circ$ at which the angular factor
passes through zero, and this is one of the reasons why 3-body correlation is expected to be so small
in ice Ih. (The other reason is that one side of the triangle is very long.)

To check whether our fitted value of $C_9$ is reasonable, we use the approximate relation
between $C_9$ and the 2-body $C_6$ dispersion coefficient due to Tang~\cite{tang69}, according
to which $C_9 \simeq \frac{3}{4} \alpha C_6$, where $\alpha$ is the molecular polarizability. (Needless
to say, all these quantities are tensors, but since the polarizability of H$_2$O is nearly isotropic we commit
only small errors by treating them as scalars.) Using the values $\alpha = 9.9$~a.u. and
$C_6 = 46$~a.u.~\cite{korona06}, we estimate $C_9 \simeq 342$~a.u., which is fairly close to our fitted value of
$289$~a.u.

The values of the 3-body CCSD(T) corrections $E^{(3)} ( \delta \mathrm{CCSD(T)} )$ to the binding
energies of the ice Ih and II structures given in Table~\ref{tab:ice} of the main text were estimated using
the ATM formula.


\clearpage


\begin{table}[!htb]
\begin{center}
\begin{tabular}{lcccc}
\hline
   & prism & cage & book & ring \\
\hline
HF({\sc molpro}) & $-41.60$ & $-41.80$ & $-43.70$ & $-45.65$ \\
HF({\sc vasp}) & $-41.44$ & $-41.92$ & $-43.65$ & $-45.74$ \\
$\Delta$MP2(direct) & $-32.14$ & $-31.85$ & $-29.53$ & $-26.25$ \\
$\Delta$MP2(EMBE-2) & $-32.21$ & $-31.85$ & $-29.35$ & $-25.94$ \\
rel MP2(EMBE-2) & $0.00$ & $0.15$ & $0.76$ & $2.22$ \\
rel MP2(ref.~\cite{bates09}) & $0.00$ & $0.10$ & $0.53$ & $1.93$ \\
$\delta \mathrm{CCSD(T)}$(direct) & $0.27$ & $0.58$ & $0.88$ & $1.13$ \\
$\delta \mathrm{CCSD(T)}$(EMBE-2) & $-0.24$ & $0.07$ & $0.63$ & $1.07$ \\
\hline
\end{tabular}
\end{center}
\caption{Components of the total binding energies (m$E_{\rm h}$ units) of four isomers 
of the H$_2$O hexamer computed
with MP2 and CCSD(T) near the complete basis-set limit. HF({\sc molpro}) and HF({\sc vasp}) are Hartree-Fock
binding energies from {\sc molpro} and from PAW calculations with {\sc vasp}, and $\Delta$MP2 
is the correlation part of total binding energy, with $\Delta$MP2(direct) computed directly from MP2
calculations on the entire cluster and $\Delta$MP2(EMBE-2) computed using the embedded MBE 
truncated at the 2-body level. Rel~MP2 indicates values of MP2 total binding energies relative
to that of the prism. Coupled-cluster corrections to the total binding energies $\delta \mathrm{CCSD(T)}$
are differences between binding energies from CCSD(T) and MP2.}
\label{tab:hexamer}
\end{table}


\begin{table}[!htb]
\begin{center}
\begin{tabular}{lccc}
\hline
   & 8-mer & 12-mer & 16-mer  \\
\hline
HF({\sc molpro}) & $-68.35$ & $-107.85$ & $-145.12$  \\
HF(\sc{vasp}) & $-68.00$ & $-107.32$ & $-144.73$ \\
HF(Wang) & $-$ & $-$ & $-144.65$ \\
$\Delta$MP2(direct) & $-41.31$ & $-$ & $-$ \\
$\Delta$MP2(EMBE-2) & $-41.28$ & $-68.94$ & $-114.95$ \\
MP2(direct) & $-109.66$ &  $-$ & $-$   \\
MP2(EMBE-2) & $-109.63$ & $-176.79$ & $-260.07$ \\
MP2(Wang) & $-$ & $-$ & $-261.57$ \\
\hline
\end{tabular}
\end{center}
\caption{Total binding energies (m$E_{\rm h}$ units) of near-equilibrium configurations of the H$_2$O
8-mer, 12-mer and 16-mer computed with MP2 near the complete basis-set limit. HF({\sc molpro})
and HF({\sc vasp}) are Hartree-Fock component of the binding energy from {\sc molpro} calculations
and from PAW calculations with {\sc vasp}; $\Delta$MP2(direct) and $\Delta$MP2(EMBE-2)
are correlation energies calculated directly with MP2 on the entire cluster and with MP2 EMBE-2 calculations.
MP2(direct) and MP2(EMBE-2)  result from addition of  $\Delta$MP2(direct) and $\Delta$MP2(EMBE-2) to
HF({\sc molpro}). HF(Wang) and MP2(Wang) values are from benchmark calculations 
of Wang \emph{et al.}~\cite{wang13}, adjusted for different choice of reference geometry of isolated monomer.
}
\label{tab:8-12-16mer}
\end{table}


\begin{table}[!htb]
\begin{center}
\begin{tabular}{lccc}
\hline
   & Ih & II & VIII  \\
\hline
HF({\sc vasp}) & $-10.40$ & $-10.07$ & $-6.84$ \\
$\Delta$MP2(EMBE-2) & $-11.97$ & $-12.17$ & $-14.60$  \\
MP2 & $-22.37$ & $-22.24$ & $-21.44$ \\
$\delta$CCSD(T)(EMBE-2) & $0.27$  & $-0.16$  & $-0.87$  \\
3-body $\delta$CCSD(T) & $0.01$ & $0.30$ & $1.22$ \\
CCSD(T) & $-22.09$  & $-22.10$ & $-21.09$  \\
\hline
$E ( \mathrm{DMC} )$ & $-22.23 \pm 0.2$ & $-22.38 \pm 0.2$ & $-21.13 \pm 0.2$ \\
$E ( \mathrm{expt} )$ & $-22.42$ & $-22.38$ & $-21.20$  \\
\hline
\end{tabular}
\end{center}
\caption{Binding energies per monomer (m$E_{\rm h}$ units) 
of the ice Ih, II and VIII structures at their
equilibrium volumes. HF({\sc vasp}) is Hartree-Fock part of binding energy computed with PAW using the {\sc vasp}
code, $\Delta \mathrm{MP2}$(EMBE-2) is MP2 correlation energy from EMBE-2 technique,
MP2 is sum of HF({\sc vasp}) and $\Delta \mathrm{MP2}$(EMBE-2), $\delta \mathrm{CCSD(T)}$ (EMBE-2) is coupled-cluster
correction from EMBE-2, 3-body $\delta$CCSD(T) is the additional
3-body CCSD(T) correction, and CCSD(T) is the sum of MP2 binding energy and coupled-cluster
corrections. DMC is benchmark binding energy from
quantum Monte Carlo~\cite{santra11}, and experimental values are from Ref.~\cite{whalley84}, with zero-point vibrational energies subtracted.}
\label{tab:ice}
\end{table}


\begin{table}[!htb]
\begin{center}
\begin{tabular}{l|ccc}
\hline
group & $n_{\rm G}$ & $E^{(3)} ( \delta \mathrm{CCSD(T)} )$ (m$E_{\rm h}$) & ATM (m$E_{\rm h}$) \\
\hline
G1& $12$ & $1.01$ & $1.01$ \\
G2 & $6$ & $0.01$ & $0.04$ \\
G3 & $6$ & $0.09$ & $0.04$ \\
G4 & $4$ & $-0.07$ & $-0.08$ \\
G5 & $24$ & $0.22$ & $0.22$ \\
G6 & $32$ & $0.07$ & $0.05$ \\
\hline
Total & $84$ & $1.22$ & $1.17$ \\
\hline
\end{tabular}
\end{center}
\caption{Three-body correlation energies contributed by different groups of trimers in ice VIII
($n_{\rm G}$ is number of trimers in each group),
calculated as the difference $E^{(3)} ( \delta \mathrm{CCSD(T)} )$ between CCSD(T)
and MP2 values of trimer 3-body energy. Also given is the 3-body correlation energy
predicted by the Axilrod-Teller-Muto formula with coefficient $C_9 = 289$~a.u.}
\label{tab:3Bcorr}
\end{table}


\clearpage

\clearpage

\clearpage

\clearpage

\clearpage

\end{document}